\documentclass[aps,prd,preprint,a4,nofootinbib]{revtex4-1}

\usepackage{xcolor}
\usepackage[colorlinks=true, pdfstartview=FitV,linkcolor=blue, citecolor=blue, urlcolor=blue]{hyperref}

\usepackage[utf8]{inputenc}
\usepackage{graphicx}
\usepackage{bm}
\usepackage{amsmath,amssymb,amsfonts,mathrsfs}
\usepackage{latexsym}
\usepackage{color}
\usepackage{tabularx}

\newcommand{\al}{\alpha}

\newcommand{\bee}{\begin{equation}}
\newcommand{\eee}{\end{equation}}

\begin{document}

\title{Stress-energy of the quantized fields in the spacetime of the Damour-Solodukhin wormhole.}

\author{Jerzy Matyjasek}
\email{jurek@kft.umcs.lublin.pl, jirinek@gmail.com}
\affiliation{Institute of Physics,
Maria Curie-Sk\l odowska University\\
pl. Marii Curie-Sk\l odowskiej 1,
20-031 Lublin, Poland}

\begin{abstract}
 The static traversable wormhole should made out of some type of exotic matter which 
 satisfies the Morris-Thorne conditions. Although the characteristic size of the 
 region with the exotic matter can be made arbitrary
 small, the calculations performed so far suggest
 that the Morris-Thorne conditions are quite restrictive and it is hard to find the matter
 with the desired properties. Traditionally, the quantized fields 
 are considered as the best candidates because they can violate the weak-energy condition.
 In this paper we employ the Schwinger-DeWitt expansion to construct and examine the approximate stress-energy tensor of the quantized 
 massive scalar (with an arbitrary curvature coupling), spinor and vector field in the spacetime
 of the Damour-Solodukhin wormhole. 
 We find that for the scalar field there is a region 
in a parameter space in which the stress-energy tensor has the desired properties. 
That means that of the twenty-one cases considered so far  (the seven types 
of the wormhole geometries and the three types of the massive fields) only in the four
cases (for certain values of the parameters) the stress-energy tensor does satisfy the Morris-Thorne
conditions.

\end{abstract}
\maketitle

\section{Introduction}
As is well known, the reaction of a black hole to small perturbations is described by a set of oscillations characterized by complex frequences.

Typically, in order to construct the traversable wormhole some amount of the strange matter is needed. 
This assertion follows from the analysis of the Einstein field equations for the general static line 
element describing a spherically-symmetric wormhole~\cite{Morris1,Morris2} (see also~\cite{Bronnikov,Ellis})
\begin{equation}
 ds^{2} = - e^{2 \Phi(r)} dt^{2} +\left(1-\frac{b(r)}{r}\right)^{-1} dr^{2} +
 r^{2} d\theta^{2} + r^{2} \sin^{2}\theta \,d\phi^{2},
 \end{equation}
where $\Phi(r)$ is a redshift function and $b(r)$ is a shape function.
In the static orthonormal frame the Einstein field equations assume simple form
\begin{equation}
 \rho = T_{\hat{t}\hat{t}} = \frac{1}{8\pi r^{2}} \frac{d}{dr}b,
\end{equation}
\begin{equation}
 \tau =  -T_{\hat{r}\hat{r}} = \frac{1}{8 \pi}\left[ \frac{b}{r^{3}} 
 - \frac{2}{r} \left( 1-\frac{b}{r}\right) \frac{d}{dr} \Phi\right]
\end{equation}
and
\begin{equation}
 p =  T_{\hat{\varphi}\hat{\varphi}} = 
T_{\hat{\theta}\hat{\theta}} = \frac{r}{2}\left[( \rho-\tau) \frac{d}{dr}\phi-\frac{d}{dr} \tau \right],
\end{equation}
for the energy density, the tension and the lateral pressure, respectively.
Now, denote by $\rho_{0}$ and $\tau_{0}$ the energy density and the tension 
at the throat. As is well known, the traversability of a  wormhole requires that
the  following minimal set of  conditions is satisfied:
\begin{equation}
 \frac{\tau_{0} -\rho_{0}}{|\rho_{0}|} \geq 0.
 \label{var2}
\end{equation}
and
\begin{equation}
 \tau_{0} >0.
 \label{var1}
\end{equation}
The second condition follows from the absence of the event horizon and  finiteness 
of the energy density $\rho_{0}$ whereas the first one is essentially 
the flaring-out condition~\cite{Morris1,Morris2}. The problem with the above 
conditions is that it is not an easy task to find matter fields satisfying them. 
Because of that the forms of matter satisfying the traversability conditions are 
called exotic. Since the quantized fields frequently violate the energy conditions 
they  are among the most solid candidates for the exotic mater. In general, to 
construct such a wormhole one has to solve the semiclassical Einstein field 
equations with the stress-energy tensor of the quantized field. The first 
(numerical) calculations of this type were carried out by Hochberg, Popov and 
Sushkov in Ref.~\cite{Hochberg}. It is a very important result showing 
that it is possible to construct the self-consistent solution of the semiclassical
Einstein field equations describing a Lorentzian wormhole connecting two
asymptotically flat regions. Unfortunately, being numerical, this approach 
does not give much information about the geometry of the thus obtained wormhole. 
On the other hand however, it seems to be extremely hard, if not impossible, to 
construct the analytic solution. The main difficulty arises because
the right hand side of the semiclassical Einstein equations should depend
functionally  on a general metric tensor or, at least, on a wide class of metrics.  
Moreover, of the all approximations of the stress-energy tensor only the one based on
the Schwinger-DeWitt expansion has the desired properties. 
The second difficulty is related to the fact that 
the stress-energy tensor is very complicated and involves higher derivatives of the metric, that
practically prevents  construction of the analytic solutions.
However, one still can gain valuable
information from the analysis of the stress-energy tensor of the quantized fields.
Specifically, one can deduce if the energy density and the tension 
have the appropriate form to support a traversable wormhole spacetime.
If the Morris-Thorne conditions are not satisfied the quantum fields 
tend to destroy a wormhole and make it less operable. Various aspects 
of the he quantum field theory in the wormhole spacetimes have been studied, for example, 
in Refs.~\cite{Carlson,Khatsymovsky,Popov2014,Bezerra,PopovS2,Popov2001,Khu,Garattini,Gar2}.
A comprehensive discussion of general properties of wormholes can be found in
Refs.~\cite{KirillBook,VisserBook,Lobo}.

It should be emphasized once again
that the inequalities (\ref{var1}) and (\ref{var2})  are quite restrictive.
Indeed, as have been shown in Refs.~\cite{Taylor,Kocuper}, of the eighteen cases considered there (six types 
of the wormhole geometries and three types of the massive fields) only in three 
cases (for certain values of the parameters) the stress-energy tensor does satisfy the Morris-Thorne
conditions. In this paper (which can be treated as an extension of Refs.~\cite{Taylor,Kocuper}) we shall construct
the stress-energy tensor of the quantized massive scalar, spinor and vector fields 
in the spacetime of the Damour-Solodukhin wormhole~\cite{Damour} and check if the matter fields
are of sufficient ``exoticity''. We shall demonstrate that for the stress-energy tensor of the 
massive scalar field with the curvature coupling $\xi$  there is a region in the space 
of parameters in which the conditions (\ref{var1}) and (\ref{var2})  are simultaneously satisfied. Unfortunately, neither the spinor nor the vector fields are of sufficient exoticity.

The line element describing the Damour-Solodukhin wormhole is given by
\begin{equation}
 ds^{2} = -\left( f(r) +\,\omega^{2} \right) dt^{2} + \frac{dr^{2}}{f(r)} +
 r^{2} d\theta^{2} + r^{2} \sin^{2}\theta \,d\phi^{2},
 \label{worm1}
\end{equation}
where $f(r) = 1-\frac{2 M}{r}$ and $\,\omega$ is a dimensionless parameter.
Although it looks as a small modification of the Schwarzschild line element, it leads to a completely
different type of the spacetime. Indeed, when $\,\omega = 0$ the line element reduces to the Schwarzschild solution, 
however, for a nonvanishing $\,\omega,$ no matter how small, it describes the Lorentzian wormhole 
with a throat located at $r =r_{0} =2 M.$ The throat joins two asymptotically flat regions
$2 M \leq r < \infty$. If the parameter $\,\omega$ is small, one expects that the  wormhole
can mimic some of the observational features of black holes. It has been shown that
for exponentially small values of the parameter, say, $\,\omega \sim e^{-4 \pi M^{2}}$ the solution (\ref{worm1})
is able to mimic both classical and quantum properties of the Schwarzschild black holes.
However, since in this paper we are less interested in the black hole mimickers we treat $\,\omega$ 
as a free parameter. 

Throughout the paper we use natural units $\hbar = c = G =1.$ The signature of the metric is taken to be ``mainly positive'' and 
the conventions for the curvature tensor are $ \mathcal{R}^{a}_{\, b c d}  = \partial_{c} \Gamma^{a}_{\,b d} ...,$ and 
$\mathcal{R}^{a}_{\, b a c} = \mathcal{R}_{bd}$.

\section{The stress-energy tensor of the quantized massive fields.}

The mathematical difficulties arising in the attempts to construct the renormalized
stress-energy tensor of the quantized fields in a curved background are well known.
They are related to the fact that the building blocks of the stress-energy tensor 
are the operator-valued distributions and the whole procedure is infected with 
unavoidable divergences. Moreover, the solutions of the equations
describing quantum fields in curved spacetime cannot be expressed, except very rare cases,  
in terms of the known special functions.  One is forced, therefore, to refer either to
approximate methods or to numerics.  The renormalized effective action of the quantized 
field in a curved background is a nonlocal quantity. However, if the Compton length 
associated with a quantized massive field is much smaller than a characteristic 
radius of curvature, then the nonlocal contribution to the total action can be neglected 
and the remaining (local) part  can be expressed as the sum of terms constructed from 
the integrated Hadamard-Minakshisundaram-DeWitt-Seely coefficients. This is the Schwinger-DeWitt
expansion in the powers of $m^{-2}.$  Once the renormalized effective action is known the stress-energy
tensor can be calculated in a standard way~\cite{FroZel2,FroZel3,ja1,ja2}. 

In a static and spherically-symmetric
background there is a complementary method of constructing the stress-energy tensor 
of the massive scalar field with an arbitrary curvature coupling devised by Anderson et al~\cite{AHS},
in which the approximate WKB solutions  of the scalar field equations are summed by means 
of the Abel-Plana formula. Both methods give precisely the same results as there 
is a one-to-one correspondence between them at each order of the calculations.
Indeed, to obtain the main approximation (i.e., $m^{-2}$ terms) the sixth-order WKB
approximation of the mode functions is needed. 
Similarly, the next-to-leading approximation (i.e., $m^{-4}$ terms) is equivalent
to the eight-order WKB approximation, and so forth.

The Schwinger-DeWitt approach has been employed in various contexts in 
Refs.~\cite{Kofman2,Kofman,Hiscock,TaylorA,jm5,jm4,jm3,jm2,jm1,Piedra} and the applications span from 
black hole physics to cosmology and form topological structures to wormholes.
Detailed analysis carried out in Refs.~\cite{AHS,Taylor} shows that the Schwinger-DeWitt approximation
of the renormalized stress-energy tensor is very good. For example, for the quantized scalar field
in the Reissner-Nordstr\"om spacetime characterized by the  mass $M$ and the charge $Q,$
the deviation of the approximate stress-energy tensor 
form the exact (numerical) one is  always below 1\%, provided the condition $m M \geq 2$ is satisified.

The approximate one-loop effective action of the quantized massive scalar, 
spinor 
and vector fields  is given by~\cite{ivanBook, Avramidi}
\begin{eqnarray}
W^{(1)}_{(s)}\,&=&\,{\frac{1}{192 \pi^{2} m^{2}}} \int d^{4}x \,g^{1/2} \left[
\alpha^{(s)}_{1} \mathcal{R}\Box \mathcal{R}\,
+\,\alpha^{(s)}_{2} \mathcal{R}_{p q} \Box \mathcal{R}^{p q}\,
+\, \alpha^{(s)}_{3} \mathcal{R}^{3} \right.  \nonumber \\
&+& \alpha^{(s)}_{4} \mathcal{R} \mathcal{R}_{p q } \mathcal{R}^{p q} 
+\,\alpha^{(s)}_{5} \mathcal{R} \mathcal{R}_{p q a b} \mathcal{R}^{p q a b}\,
+\, \alpha^{(s)}_{6} \mathcal{R}^{p}_{q} \mathcal{R}^{q}_{a} \mathcal{R}^{a}_{p} 
+ \alpha^{(s)}_{7}\mathcal{R}^{p q} \mathcal{R}_{a b} \mathcal{R}^{a ~ b}_{~ p ~ q}  \nonumber \\
&+&\left. \, 
\,\alpha^{(s)}_{8} \mathcal{R}_{p q} \mathcal{R}_{~ c a b}^{p} \mathcal{R}^{q c a b}\,
+\, \alpha^{(s)}_{9} {\mathcal{R}_{a b}}^{p q} {\mathcal{R}_{p q}}^{c d} {\mathcal{R}_{ c d}}^{a b} 
+ \alpha^{(s)}_{10} \mathcal{R}^{a ~ b}_{~ p ~ q} \mathcal{R}^{p ~ q}_{~ c ~ d} \mathcal{R}^{c ~d}_{~ a ~ b}\right]  
\nonumber \\
&=&{\frac{1}{192 \pi^{2} m^{2}}}\sum_{i = 1}^{10} \alpha^{(s)}_{i} \mathcal{W}_{i},
\label{action}
\end{eqnarray}
where $m$ is the mass of the field  and $\mathcal{W}_{i}$ are purely
geometric terms constructed entirely form the Riemann tensor, its contractions and 
covariant derivatives. The type of the field is encoded in the coefficients $\alpha_{i}^{(s)}$ 
tabulated in Table~\ref{table1}.
\begin{table}[h]
\caption{The coefficients $\al_{i}^{(s)}$ for the massive scalar,  spinor, and 
vector fields}
\begin{tabular}{|c|c|c|c|}
\toprule
&s=0&s=1/2&s=1\\ \hline
$\al^{(s)}_{1}$&$\frac{1}{2}\xi^{2}\,-\,\frac{1}{5}\xi$\,+\,$\frac{1}{56}
$&$-\frac{3}{140}$&$-\frac{27}{280}$\\	
$\al^{(s)}_{2}$&$\frac{1}{140}$&$\frac{1}{14}$&$\frac{9}{28}$\\$\al^{(s)}_{3}
$&$\left(\frac{1}{6}-\xi\right)^{3}$&$\frac{1}{432}$&$-\frac{5}{72}$\\
$\al^{(s)}_{4}$&$-\frac{1}{30}\left(\frac{1}{6}-\xi\right)$&$-\frac{1}{90}
$&$\frac{31}{60}$\\$\al^{(s)}_{5}$&$\frac{1}{30}\left(\frac{1}{6}-\xi\right)$&$
-\frac{7}{720}$&$-\frac{1}{10}$\\$\al^{(s)}_{6}$&$-\frac{8}{945}$&$-\frac{25}{
378}$&$-\frac{52}{63}$\\$\al^{(s)}_{7}$&$\frac{2}{315}$&$\frac{47}{630}$&$-\frac
{19}{105}$\\	
$\al^{(s)}_{8}$&$\frac{1}{1260}$&$\frac{19}{630}$&$\frac{61}{140}$\\	
$\al^{(s)}_{9}$&$\frac{17}{7560}$&$\frac{29}{3780}$&$-\frac{67}{2520}$\\	
$\al^{(s)}_{10}$&$-\frac{1}{270}$&$-\frac{1}{54}$&$\frac{1}{18}$\\ \botrule	
\end{tabular}
\label{table1}
\end{table}

The renormalized stress-energy tensor is given by 
\begin{equation}
\langle T^{a b}\rangle  =
\frac{2}{g^{1/2}}\frac{\delta }{\delta g_{ab}} W_{(s)}^{(1)} = \frac{1}{96\pi^{2}  m^{2}{g^{1/2}}}    
\sum_{i=1}^{10} \alpha_{i}^{(s)} \frac{\delta}{\delta g_{ab}} \mathcal{W}_{i}
\label{gen_ten}
\end{equation}
and the full form of the functional derivatives of $\mathcal{W}_{i}$ with respect to the 
metric tensor was given in Refs.~\cite{ja1,ja2}. The final result is quite complicated and 
will not be presented here. 

Because of the simplicity of the line element describing static and 
spherically-symmetric line element one can use
the Euler-Lagrange equations to construct the $(00)$ and $(11)$ component of the stress-energy tensor.
The angular components can be calculated from the covariant conservation equation $\nabla_{a} T^{ab} =0.$
Although in this method one avoids calculations of the functional derivatives of the general action with respect
to the metric tensor $g_{ab}$, the result expressed in terms of the two metric potential functions 
are still too complicated to be presented here.

Now, let us return to the Damour-Solodukhin wormhole. Although the  line element describing 
the Damour-Solodukhin wormhole is a small modification of the Schwarzschild metric tensor,
the stress-energy tensor of the quantized massive fields calculated with the aid of 
Eqs.~(\ref{action}) and (\ref{gen_ten}) is quite complicated and for obvious reasons 
will not be given here\footnote{The components of the stress-energy tensor can be obtained 
on request from the author. }. Instead, we will focus on its behavior at the throat and analyze if 
the stress-energy tensor can support the existence of the Damour-Solodukhin wormhole.

After some algebra, the stress energy of the massive scalar field with a general curvature coupling
$\xi$ can be written in the form
\begin{equation}
 T_{a}^{b} = \frac{1}{96 \pi^{2} m^{2} M^{6}} \left(1-\frac{2}{x} -\,\omega^{2} \right)^{-6} \sum_{k=0}^{7} \beta_{a }^{(k) b} \frac{1}{x^{k+8}},
 \label{q_gen}
\end{equation}
where $x = r/M$ and the numerical coefficients $\beta_{a}^{(k) b}$  depend parametrically on $\,\omega$ and $\xi.$ 
By construction it is covariantly conserved and regular.
It can be easily shown
that when $\,\omega =0$ the tensor reduces to the known result in the Schwarzschild spacetime. 
On the other hand, at the throat of the Damour-Solodukhin wormhole
one has
\begin{equation}
 T_{a}^{b} = \frac{1}{96 \pi^{2} m^{2} M^{6} \,\omega^{6}}\, \mathfrak{T}_{a}^{b},
\end{equation}
where the tensors $\mathfrak{T}_{a}^{b}$ are given by
\begin{eqnarray}
 \mathfrak{T}_{t}^{t} &=&
\frac{43\, \,\omega^6}{35840}+\frac{907\, \,\omega^4}{215040}+\frac{23 \,\omega^2}{4480} +\frac{179}{107520}  
-\left(\frac{3 \,\omega^2}{64}+\frac{13}{512}\right) \xi^3
\nonumber \\
&&
+\left(\frac{13 \,\omega^4}{128}+\frac{33 \,\omega^2}{256}+\frac{47}{1024}\right) \xi ^2
-\left(\frac{9 \,\omega^6}{1280}+\frac{59 \,\omega^4}{1536}+\frac{3 \,\omega^2}{64}+\frac{81}{5120}\right) \xi ,
\end{eqnarray}
\begin{eqnarray}
 \mathfrak{T}_{r}^{r} &=& \frac{23 \,\omega^6}{35840}+\frac{131 \,\omega^4}{215040}+\frac{\,\omega^2}{1920}
 + \frac{11}{107520}-\left(\frac{3 \,\omega^2}{256}+\frac{1}{512}\right) \xi^3
\nonumber \\
&&
+\left(\frac{\,\omega^4}{64}+\frac{9 \,\omega^2}{512}+\frac{3}{1024}\right) \xi ^2
-\left(\frac{3 \,\omega^6}{1280}+\frac{49 \,\omega^4}{7680}+\frac{43 \,\omega^2}{7680}+\frac{1}{1024}\right) \xi 
\end{eqnarray}
and
\begin{eqnarray}
\mathfrak{T}_{\theta}^{\theta} &=& \mathfrak{T}_{\phi}^{\phi} = 
\frac{5 \,\omega^6}{3584}+\frac{7 \,\omega^4}{2048}+\frac{173 \,\omega^2}{43008} +\frac{283}{215040}  
-\left(\frac{21 \,\omega^2}{512}+\frac{23}{1024}\right) \xi^3\nonumber \\
&&
+\left(\frac{3 \,\omega^4}{32}+\frac{121 \,\omega^2}{1024}+\frac{87}{2048}\right) \xi ^2
-\left(\frac{3 \,\omega^6}{512}+\frac{19 \,\omega^4}{512}+\frac{659 \,\omega^2}{15360}+\frac{147}{10240}\right) \xi. 
\end{eqnarray}
The stress-energy tensor of the massive spinor field is still of the form (\ref{q_gen})
with the coefficients $\beta_{a}^{(k)b}$ depending solely on $\,\omega.$
The components of $\mathfrak{T}_{a}^{b}$ tensor can be written in the form
\begin{equation}
 \mathfrak{T}_{t}^{t}= -\frac{3 \,\omega^6}{3584}+\frac{233 \,\omega^4}{71680}+\frac{47 \,\omega^2}{17920}+\frac{269}{430080},
\end{equation}
\begin{equation}
\mathfrak{T}_{r}^{r}= \frac{3 \,\omega^6}{4480}-\frac{19 \,\omega^4}{107520}+\frac{\,\omega^2}{15360}+\frac{17}{430080}
\end{equation}
and
\begin{equation}
\mathfrak{T}_{\theta}^{\theta} = \mathfrak{T}_{\phi}^{\phi} = \frac{33 \,\omega^6}{35840}-\frac{11 \,\omega^4}{10240}-\frac{317 \,\omega^2}{430080}-\frac{143}{860160}.
\end{equation} 
Similarly, for the components of the stress-energy tensor of the massive
vector field, one has
\begin{equation}
  \mathfrak{T}_{t}^{t} = -\frac{53 \,\omega^6}{35840}+\frac{7019 \,\omega^4}{215040}+\frac{201 \
\,\omega^2}{8960}+\frac{179}{35840},
\end{equation}
\begin{equation}
\mathfrak{T}_{r}^{r}=  \frac{83 \,\omega^6}{35840}-\frac{181 \,\omega^4}{215040}+\frac{\,\omega^2}{1280}+\frac{11}{35840}
\end{equation}
and
\begin{equation}
 \mathfrak{T}_{\theta}^{\theta} = \mathfrak{T}_{\phi}^{\phi}=\frac{19 \,\omega^6}{17920}-\frac{51 \,\omega^4}{10240}-\frac{79 \,\omega^2}{14336}-\frac{19}{14336}.
\end{equation}

Before starting our discussion of the Morris-Thorne conditions let us count the general properties of the obtained stress-energy tensors. 
By construction, they are regular and covariantly conserved. Inspection of (\ref{q_gen}) shows that the components of  $T_{a}^{b}$ change sign a few times  in the region close to $r_{0}$ and  rapidly go to 0  as $r \to \infty.$ The energy density at the throat (and by continuity in its vicinity)
is negative for $\omega^{2} <4.6$ and $\omega^{2} < 22.7458$ for the spinor and the vector field, respectively.
A different picture emerges form the analysis of the minimally coupled and the conformally coupled scalar fields.
Indeed, in the both cases the energy density is always negative for $\omega^{2} >0.$ In summary, 
in  all considered cases there is a pocket of negative energy in the vicinity of the throat
for small values of $\omega$ (say $\omega^{2} < 1$).

\section{Discussion}

Although the stress-energy tensor of the quantized fields is interesting in its own right, now we shall 
examine whether the quantum fields it describes are of sufficient exoticity to support the Damour-Solodukhin wormhole. 
In our context, the energy density and the radial tension are defined as
\begin{equation}
 \rho = -T_{t}^{t}
\end{equation}
and 
\begin{equation}
\tau = -T_{r}^{r}.         
\end{equation}
In what follows we shall not impose any additional restrictions on the parameter $\,\omega.$

Let us start with the massive spinor fields. Inspection of the components of the stress-energy tensor at
the throat of the wormhole
shows that the first Morris-Thorne condition~(\ref{var1}) is satisfied for $\,\omega < 1.7044.$ On  the other hand,
the radial tension is always negative
and consequently the components of the stress-energy of the massive spinor field will not help to support the Damour-Solodukhin wormhole.
Similarly, the first condition for the massive vector field is satisfied for $\,\omega < 3.0727$ 
whereas the radial tension is always negative. Since the second condition cannot be satisfied, the stress-energy
tensor does not have the form required to support the wormhole.

\begin{figure}
\centering
\includegraphics[width=11cm]{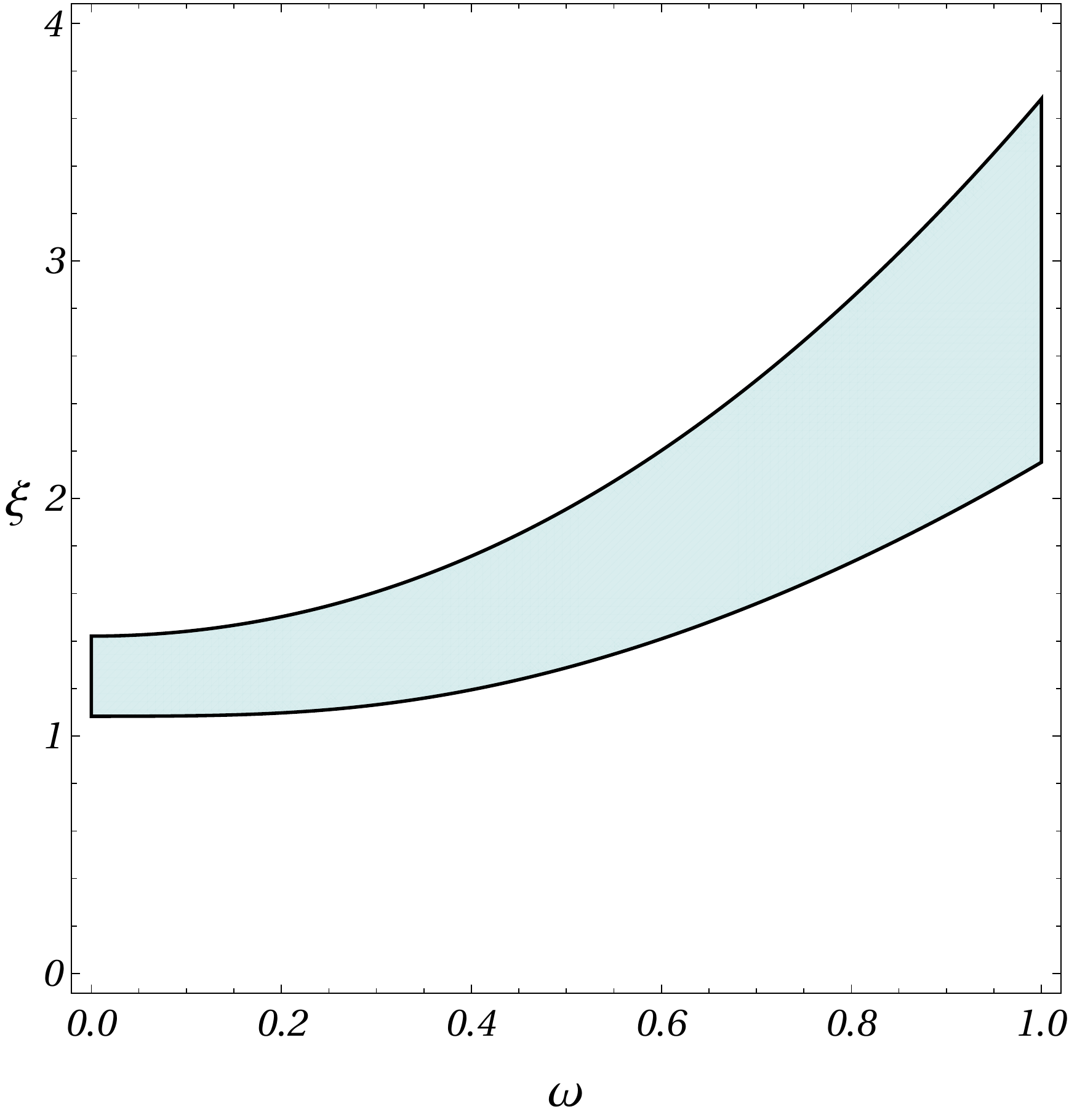}
\caption{The shaded region of the $(\,\omega,\xi)$-plane represents values  of the parameters for which
the Morris-Thorne conditions are satisfied. Both conformally and minimally coupled fields lie outside the
shaded region.}
\label{fig4}
\end{figure}

Now, let us consider the massive scalar field with the general curvature coupling and examine
if the Morris-Thorne conditions can be satisfied simultaneously in the two-dimensional parameter space
$(\,\omega, \xi).$ The results of the numerical calculations are illustrated graphically. Inspection of 
Fig.\ref{fig4} shows
that there is a region in the $(\,\omega,\xi)$-plane in which the stress-energy tensor has the form required to 
support the wormhole. It is quite important result as it shows that within the framework of the 
Schwinger-DeWitt approximation it is possible to construct the stress-energy tensor with the desired properties.
Of the twenty-one cases analyzed so far (with the result of this paper included), only in the four cases the stress-energy tensor has the required form to sustain the wormhole throat.
It should be noted however, that if the physical values of the coupling parameter are chosen, 
i.e., either the  minimal, $\xi =0,$  or the confromal, $\xi=1/6,$ coupling,
then the energy density and the radial tension tend to destroy all the wormholes.  
The general picture that emerges from the analysis is that the scalar case is preferable
as it (in our formulation of the theory) allows for an arbitrary value of the coupling parameter.
Moreover, if the wormhole geometry depends on at least one parameter then the space of 
parameters can be sufficiently rich and can contain the region(s) in which the Morris-Thorne
conditions are satisfied. 

This somewhat pessimistic conclusion does not mean that the stress-energy tensor of other fields is useless in the context of 
wormholes. Indeed, since the stress-energy tensor constructed within the Schwinger-DeWitt framework
depends functionally on a general metric, one could try to solve the semiclassical Einstein field equations
self-consistently. Consider, for simplicity, the general static and spherically-symmetric spacetime described by the line element 
\begin{equation}
 ds^{2} = - F(l) dt^{2} + dl^{2} + r^{2}(l) \left( d\theta^{2} + \sin^{2} \theta d\phi^{2}  \right),
\label{popov}
\end{equation}
where $F(l)$ and $r(l)$ are two functions of the proper distance $l.$ 
For this line element the semiclassical Einstein field equations 
\begin{equation}
 R_{a}^{b}[g] - \frac{1}{2}  R[g] \delta_{a}^{b} = 8 \pi T_{a}^{b}[g]
\end{equation}
reduce to the system of the two independent ordinary differential equations. Imposing the appropriate boundary conditions one can attempt to solve the equations numerically and decide 
if the thus obtained solution describes traversable wormhole. This work is in progress and the results
will be published elsewhere.

\end{document}